\documentclass[a4paper,11pt]{article}
\pdfoutput=1
\usepackage{jheppub}

\newcommand{\mathsym}[1]{{}}

\newcommand{\ba}{\begin{array}}
\newcommand{\ea}{\end{array}}
\newcommand{\bal}{\begin{align}}
\newcommand{\eal}{\end{align}}
\newcommand{\be}{\begin{equation}}
\newcommand{\ee}{\end{equation}}
\newcommand{\beqa}{\begin{eqnarray}}
\newcommand{\eeqa}{\end{eqnarray}}

\def\321{$SU(3)\times SU(2)\times U(1)$}

\title{Searching for Flavored Gauge Bosons}
\author[a]{Eung Jin Chun,}
\author[a]{Arindam Das,}
\author[b,1]{Jinsu Kim,\note{Corresponding Author}}
\author[c]{and Jongkuk Kim}
\affiliation[a]{Korea Institute for Advanced Study, Seoul 02455, Korea}
\affiliation[b]{Institute for Theoretical Physics, Georg-August University G\"ottingen, Friedrich-Hund-Platz 1, G\"ottingen, D-37077 Germany}
\affiliation[c]{Department of Physics, BK21 Physics Research Division, Institute of Basic Science, Sungkyunkwan University, Suwon 440-746, Korea}

\emailAdd{ejchun@kias.re.kr}
\emailAdd{arindam@kias.re.kr}
\emailAdd{jinsu.kim@theorie.physik.uni-goettingen.de}
\emailAdd{jongkukkim@skku.edu}

\abstract{
Standard Model may allow an extended gauge sector with anomaly-free flavored gauge symmetries, such as $L_{i} - L_{j}$, $B_{i} - L_{j}$, and $B - 3L_{i}$, where $i,j=1,2,3$ are flavor indices.
We investigate phenomenological implications of the new flavored gauge boson $Z^{\prime}$ in the above three classes of gauge symmetries. Focusing on the gauge boson mass above 5 GeV, we use the lepton universality test in the $Z$ and $\tau/\mu$ decays, LEP searches, LHC searches, neutrino trident production bound, and LHC $Z\rightarrow 4\mu$ searches to put constraints on the $g^{\prime}-M_{Z^{\prime}}$ plane.
When $L_1$ is involved, the LEP bounds on the $e^{-}e^{+} \rightarrow \ell^{-}\ell^{+}$ processes give the most stringent bounds, while the LHC bound becomes the strongest constraints in the large $M_{Z^{\prime}}$ region when $B_{i}$ is involved. The bound from $Z\rightarrow 4\mu$ productions, which is applicable for $L_2$-involved scenarios, provides stringent bounds in the small $M_{Z^{\prime}}$ region.
One exception is the $B-3L_2$ scenario, in which case only a small region is favored due to the lepton universality.
}

\begin{document}
\maketitle
\flushbottom

\section{Introduction}
\label{sec:intro}

The gauge invariance of the Standard Model (SM) allows accidental global symmetries like $B$ and $L$ out of which $B-L$ is anomaly-free and thus could be extended to a gauge symmetry. There also appear flavor-dependent combinations which are anomaly-free such as $L_i-L_j$, $B-3L_i$, or $B_i -L_j$. These could be extended to gauge symmetries at high energy.
Of course, any linear combination of the above ``flavored symmetries'' is also anomaly-free and thus can be a gauge symmetry. 
Searches for such a new gauge boson can be carried out in the vast ranges of the mass and coupling. In particular, a light gauge boson below around 10 GeV are known to be highly constrained by collections of low-energy experiments in generic dark photon models \cite{dark16,Ilten:2018crw} or $L_i-L_j$ gauge symmetry \cite{bauer18}. Such studies can be extended to the above flavored gauge bosons.
In this paper, focusing on the mass range above 5 GeV, we aim to investigate phenomenological implications of flavored gauge bosons and various limits from existing searches.

The new gauge interaction behaves like the SM $Z$ interaction wherever applicable, and thus its contribution to SM observables should be suppressed well below the standard $Z$ contribution. 
Being flavor-dependent, it can also lead to sizable modification to the flavor-universal SM interaction, and thus some of significant limits come from the lepton universality in the $Z$ and $\tau/\mu$ decays. When $L_1$ is involved, the extra gauge boson $Z'$ can be produced at LEP, providing the most stringent bound. Likewise, the LHC search can be applied when $Z'$ couples to the $B_i$ current. See Refs.~\cite{Ilten:2018crw,Babu:2017olk,Basso:2008iv,Accomando:2016rpc,Batell:2016zod} for previous studies.
Observations of the neutrino trident production and $Z \rightarrow 4\mu$ at the LHC give extra constraints when $L_2$ is involved.
See Refs.~\cite{Altmannshofer:2014cfa,Altmannshofer:2016jzy} for the $L_{2}-L_{3}$ scenario.

The paper is organized as follows. In Section \ref{sec:model} we introduce three different classes of flavored gauge symmetries. We then test the lepton universality, using the SLD and LEP experimental data, and place limits on the model parameters in Section \ref{sec:LUtest}.
LEP and BaBar search are taken into account, which are appropriate for $L_1$-involved models, in Section \ref{sec:LEPbound}.
In Section \ref{sec:LHCbound} we look at the LHC bound which can be applied when $B_i$ is involved.
In Section \ref{sec:NTPand4muLHC} we discuss bounds from the neutrino trident production and $Z\rightarrow 4\mu$ at the LHC which are applicable for $L_2$-involved scenarios.
We present our results in Section \ref{sec:result} and conclusions in Section \ref{sec:con}. All the analytical expressions are summarized in Appendices.

\section{Flavored gauge interactions}
\label{sec:model}

Let us consider flavored gauge symmetries under which the left-handed and right-handed fermions (quarks and leptons) transform equivalently, and thus the SM gauge invariance extended to $SU(2)_L \times SU(2)_R$ is respected. 
For the $L_i - L_j$ gauge invariance \cite{LiLj}, we have the interaction Lagrangian,
\begin{align}
  \mathcal{L}^{L_i-L_j}_{{\rm int}}
  =
  -g^{\prime}Z^{\prime\mu}\left(
  \bar{\ell}_{i}\gamma_{\mu}\ell_{i}
  -\bar{\ell}_{j}\gamma_{\mu}\ell_{j}
  \right)\,,
\end{align}
where $\ell_i$ contains the $i$-th generation of the $SU(2)_L$ doublet $(\nu_L, e_L)$ and $SU(2)_R$ doublet $(\nu_R, e_R)$.
The latter includes the right-handed neutrino $\nu_R$ which can get a Majorana mass after the $L_i-L_j$ symmetry breaking.
A typical difficulty with such a flavored gauge symmetry is generation of the observed lepton (quark) mixing, which may require a judicial choice of the Higgs sector \cite{heeck11,Babu:2017olk}.

Similarly, one can write down the $B-3L_{i}$ gauge interaction Lagrangian,
\begin{align}
  \mathcal{L}^{B-3L_{i}}_{{\rm int}}
  =
  -g^{\prime}Z^{\prime\mu}\left( \frac{1}{3}
  \sum_{j=1}^3 \bar{q}_{j}\gamma_{\mu}q_{j}
  -3 \bar{\ell}_{i}\gamma_{\mu}\ell_{i}
  \right)\,,
\end{align}
where $q_j$ represents the $j$-th generation of the left- and right-handed quark doublet.
We also consider the $B_i -L_j$ gauge invariance with the interaction Lagrangian,
\begin{align}
  \mathcal{L}^{B_i -L_j}_{{\rm int}}
  =
  -g^{\prime}Z^{\prime\mu}\left(
  \frac{1}{3}\bar{q}_{i}\gamma_{\mu}q_{i}
  -\bar{\ell}_{j}\gamma_{\mu}\ell_{j}
  \right)\,.
\end{align}
Note that gauging flavored baryon number requires a nontrivial Higgs sector to generate a viable CKM matrix for the quark sector.

Any linear combination of those three flavored gauge symmetries can also be a gauge symmetry. Focusing on the above three classes, we apply the lepton universality tests, LEP and BaBar bound, LHC bound, and bounds from neutrino trident and $Z\rightarrow 4\mu$ productions.

We end this section with the decay width of the process $Z^{\prime} \to f \bar f$ for each fermion:
\begin{align}
	\Gamma\left( Z^{\prime} \to f \bar{f} \right) 
	= 
	\frac{N_c Q_f'^2 g^{\prime 2}}{12\pi  M_{Z^{\prime}}}  \left( M^2_{Z^{\prime}} +2m^2_f \right)
	\sqrt{1-\frac{4m^2_f}{ M^2_{Z^{\prime}}}}
	\theta\left(M_{Z^{\prime}} -2m_f \right)\,,
\end{align}
where $\theta$ is the step function, $f$ is either lepton or quark, $N_c=3$ $(1)$ for quark (lepton), and $Q'$ is the flavored gauge charge of $f$.

\section{Lepton universality tests}
\label{sec:LUtest}

Precision electroweak tests have been performed to establish lepton universality at the level of $0.1\%$. 
Lepton-flavored gauge interactions act non-universally and thus induce sizable deviations to the lepton universality at the loop level which can be constrained by such precision measurements. 

First, let us consider the lepton universality test taken by the SLD and LEP experiments with data taken at the Z resonance \cite{Zll}.
The measurements in $Z$ decays lead to the following ratios of the leptonic branching fractions:
\begin{align}\label{lu-zdecay}
	{\Gamma_{Z\to \mu^+ \mu^-}\over \Gamma_{Z\to e^+ e^- }} = 1.0009 \pm 0.0028
	\,,\qquad
	{\Gamma_{Z\to \tau^+ \tau^- }\over \Gamma_{Z\to e^+ e^- }} = 1.0019 \pm 0.0032
	\,,
\end{align}
with a correlation of $+0.63$. For each lepton-flavored gauge interaction, we derive the following quantities:
\begin{align}\label{eqn:deltaellell}
	\delta_{\ell \ell } \equiv {\Gamma_{Z\to \ell^+ \ell^-} \over \Gamma_{Z\to e^+ e^- }}-1
\end{align}
for $\ell=\mu$ and $\tau$ at one-loop level to compare with the measurements. 
For our calculation, we use the SM value $g_L^e=-0.27$ and $g_R^e=s_W^2=0.23$ which are measured also by the electroweak precision test \cite{Zll}.
We present expressions for $\delta_{\ell\ell}$ in Appendix A.

Another lepton universality test can be made by HFAG \cite{hfag16} in the pure leptonic and semi-hadronic processes: 
$\ell \to \ell' \nu \bar \nu$, $\tau \to \pi/K \nu$, and $\pi/K\to \mu \bar \nu$.
HFAG determined the branching ratios for each process which can be translated to the ratios of flavor-dependent couplings as follows:
\begin{eqnarray} \label{hfag-data}
&&
\left( g_\tau \over g_\mu \right) = 1.0010 \pm 0.0015, \quad
\left( g_\tau \over g_e \right) = 1.0029 \pm 0.0015, \quad
\left( g_\mu \over g_e \right) = 1.0019 \pm 0.0014, 
\nonumber\\
&&
\left( g_\tau \over g_\mu \right)_\pi = 0.9961 \pm 0.0027, \quad
\left( g_\tau \over g_\mu \right)_K= 0.9860 \pm 0.0070,
\end{eqnarray}
with the correlation matrix
\begin{equation} \label{hfag-corr}
\left(
\begin{array}{ccccc}
1 & +0.53 & -0.49 & +0.24 & +0.11 \\
+0.53  & 1     &  + 0.48 & +0.26    & +0.10 \\
-0.49  & +0.48  & 1       &   +0.02 & -0.01 \\
+0.24  & +0.26  & +0.02  &     1    &     +0.06 \\
+0.11  & +0.10  & -0.01  &  +0.06  &   1 
\end{array} \right) .
\end{equation}
Following the recipe in Ref.~\cite{chun16}, we compare the experimental determination with the model prediction of
\begin{align}
	\delta_{\ell/\ell^{\prime}}
	\equiv 
	\left(
	g_{\ell} \over g_{\ell^{\prime}} 
	\right) -1
	\,, \qquad
	\delta_{\pi, K}
	\equiv 
	\left(
	g_\tau \over g_\mu
	\right)_{\pi, K} -1
	\,,
\end{align}
at one-loop level, for each flavored gauge interaction.
We summarize expressions for $\delta_{\ell/\ell^{\prime}}$ and $\delta_{\pi,K}$ in Appendix B.

\section{LEP and BaBar bound}
\label{sec:LEPbound}

LEP search can be done with $e^+ e^- \to Z' \to \ell^+ \ell^-$ or $q \bar q$ \cite{Schael:2013ita}.
We consider bounds coming from the differential cross sections of $e^+ e^- \to Z' \to \ell^+ \ell^-$ processes with $\ell = e,\mu,\tau$.
We take experimental data for differential cross sections of $e^+ e^- \to e^+ e^-$ shown in Tables 3.11 and 3.12, $e^+ e^- \to \mu^+ \mu^-$ in Table 3.8, and $e^+ e^- \to \tau^+ \tau^-$ in Table 3.9 of Ref.~\cite{Schael:2013ita}.
BaBar searches for the process $e^+ e^- \to \gamma\, Z'(\to e^+ e^- , \mu^+ \mu^-)$ \cite{babar}.
Thus the LEP and BaBar bounds are applicable to flavored gauge symmetry containing $L_1$.

Expressions for the differential cross sections for $e^{+}e^{-} \rightarrow \ell^{+}\ell^{-}$ are given in Appendix C.

\section{LHC bound}
\label{sec:LHCbound}
For the $B_{i}$-involved gauge symmetries, such as $B_{i} - L_{j}$ and $B - 3L_{i}$, LHC bounds on $pp\to Z' \to l^+ l^-$, $jj$, or $b \bar b$ \cite{Aaboud:2017buh,Sirunyan:2018exx,Aaboud:2017sjh} can be applied.

For example, in Ref.~\cite{Ilten:2018crw}, a light $Z^{\prime}$ between 0.002 GeV and 90 GeV is studied in the $B-L$ scenario. They obtained the limits by using $Z^{\prime} \rightarrow 2\mu$ final state and comparing with $B$ meson experiments.
In Ref.~\cite{Basso:2008iv}, the limits on the $Z^{\prime}$ coupling for the $B-L$ model were examined in the (500 GeV, 5 TeV) mass range. They considered decays of $Z^{\prime}$ to two right-handed neutrinos which subsequently decay into $\ell+\nu/\text{hadron}$. 
In Refs.~\cite{Accomando:2016rpc}, the Drell-Yan process was studied in the $B-L$ model. They considered heavy $Z^{\prime}$ and the LHC bound turned out to be weak.
In Ref.~\cite{Batell:2016zod} the limits were obtained by considering the pair production of right-handed neutrinos from the $Z^{\prime}$ boson for the same $B-L$ model. They considered $Z^{\prime} \rightarrow 2\mu$ modes to constrain the coupling in the (1 GeV, 500 GeV) mass range.
A $B_{3} - L_{3}$ scenario, with the addition of one extra SM-singlet scalar in the two-Higgs-doublet model, was studied in Ref.~\cite{Babu:2017olk} in the (1 MeV, 100 GeV) mass range.
In our analysis we calculate the cross sections for the dilepton production from $Z^{\prime}$ for several aforementioned flavored scenarios. We then put the limits on $g^{\prime}$ by comparing the cross sections with the existing $Z^{\prime} \rightarrow 2\ell$ searches at LHC \cite{Aaboud:2017buh,Sirunyan:2018exx,Aaboud:2017sjh}.

For the $B_i-L_j$ scenario we consider the production of the $Z^\prime$ from different generations of quarks and decays into three different generations of leptons.
We first calculate the $Z^\prime$ production cross section by considering the first generation quarks $(u, d)$ in the initial state followed by the decay into $e,$ $\mu$, and $\tau$.
We compare the cross sections with the heavy resonance $(Z^\prime)$ production at ATLAS \cite{Aaboud:2017buh}. CMS has also tested such processes for the $Z^\prime$ production \cite{Sirunyan:2018exx}.
CMS compared the $pp\to Z^\prime\to 2\ell$ with $pp \to Z \to 2\ell$ with $\ell=e, \mu$, while ATLAS considered $pp\to Z^{\prime}\to 2\ell$ process to calculate the bounds.
We therefore consider the ATLAS results for the $e$ and $\mu$ to compare directly with our scenario.

In these searches \cite{Aaboud:2017buh, Sirunyan:2018exx} different models like SSM and $Z^\prime_\psi$ are taken into consideration where the $Z^{\prime}$ decays into $e$ and $\mu$ after being produced in the proton-proton collision (See Ref.~\cite{Langacker:2008yv} for a review). Conservatively we consider these limits in our case for the $Z^\prime$ production cross section through the $e$ and $\mu$ final states and compare with the cross sections in our scenario.
For the $\tau$ case, ATLAS and CMS both have considered the $pp\to Z^\prime \to 2\tau$ process~\cite{Aaboud:2017sjh}. Thus we consider both of the ATLAS and CMS results to compare the di-tau production from $Z^\prime$ in our analysis.
We take the detector efficiencies of the $e$ and $\mu$ as $85\%$ and $95\%$ respectively at the LHC. We also consider the $\tau$ tagging efficiency as $60\%$ when $\tau$ dominantly decays hadronically at the LHC.
Note that we are considering the observed limits at the ATLAS and CMS experiments.
We analyze the $B_{2}-L_{j}$ and $B_{3}-L_{j}$ scenarios in the same way. In the latter case we take into account only the bottom quark in the production process.

In the $B-3L_{i}$ case, the coupling between $Z^\prime$ and the leptons will be three times larger than the previous case, affecting not only the production cross section but also the total decay width of the $Z^\prime$. 
We follow the same procedure as before in order to constrain $g^{\prime}$ and $M_{Z^{\prime}}$.

\section{Neutrino trident and $Z \rightarrow 4\mu$ productions}
\label{sec:NTPand4muLHC}
Observations of the neutrino trident production and $Z \rightarrow 4\mu$ at the LHC provide strong constraints on the $L_2$-involved gauge symmetries.
Various neutrino beam experiments such as CHARM-II \cite{Geiregat:1990gz} and CCFR \cite{Mishra:1991bv} have established neutrino trident production, $\nu_\mu N \to \nu_\mu N \mu^+ \mu^-$. 
The observed cross sections are as follows:
\begin{align}
	\frac{\sigma_{\rm CHARM-II}}{\sigma_{\rm SM}} = 1.58 \pm 0.57
	\,,\qquad
	\frac{\sigma_{\rm CCFR}}{\sigma_{\rm SM}} = 0.82 \pm 0.28 \,.
\end{align}
The observed scattering cross sections are consistent with the SM expectation.
Thus, the $L_2$-involved gauge models can be strongly constrained by the neutrino trident production. 
In our analysis, we take $2\sigma$ exclusion limit from the CCFR observation \cite{Altmannshofer:2014pba}.

LHC has measured $pp \rightarrow Z \rightarrow 4\mu$ channels.
In the measurements, four muon events with an invariant mass near $Z$ boson mass are selected.
In the $L_2$-involved gauge models, the produced $Z$ boson might decay to $Z'$ and a muon pair.
The produced $Z'$ subsequently decays to a pair of the SM fermions if kinematically allowed.    
Thus, the selected events are sensitive to decay processes including $Z\rightarrow Z'\mu\mu \rightarrow 4\mu$ decays.
For $M_{Z'} \geq M_b$, the branching ratios of the $Z'$ boson are $1/4$, $3/8$, and $27/59$ for $L_i - L_2$, $B_i - L_2$, and $B-3L_2$ cases, respectively. 
In our analysis, $Z'$ decays to a pair of top quarks are kinematically forbidden due to heavy top quark mass.
Using the rescaled branching ratio analysis \cite{Sirunyan:2018nnz}, we put $2\sigma$ exclusion limits. 
See also Refs.~\cite{Elahi:2015vzh} for another methods applied to the $L_2 - L_3$ model.
Similar bounds were obtained in Refs.~\cite{Altmannshofer:2014cfa,Altmannshofer:2016jzy} for the $L_{2} - L_{3}$ model. Our $Z \rightarrow 4\mu$ bound is stronger than those in Ref.~\cite{Altmannshofer:2016jzy}, while the neutrino-trident bound remains the same. For example, in Ref.~\cite{Altmannshofer:2016jzy}, the stringent bound on $g^{\prime}$ from $Z \rightarrow 4\mu$ was derived to be about 0.015 at $M_{Z^{\prime}} \simeq 10$ GeV. Our bound, for the $L_2 - L_3$ model, is $g^{\prime} \lesssim 0.004$ at $M_{Z^{\prime}} \simeq 10$ GeV. This is mainly due to the updates in experimental data.

\section{Results}
\label{sec:result}

Taking all the considerations given in previous sections into account, we put limits on the $g^\prime$ versus $M_Z^\prime$ plane.

\begin{figure*}[t]
\begin{center}
\includegraphics[scale=0.5]{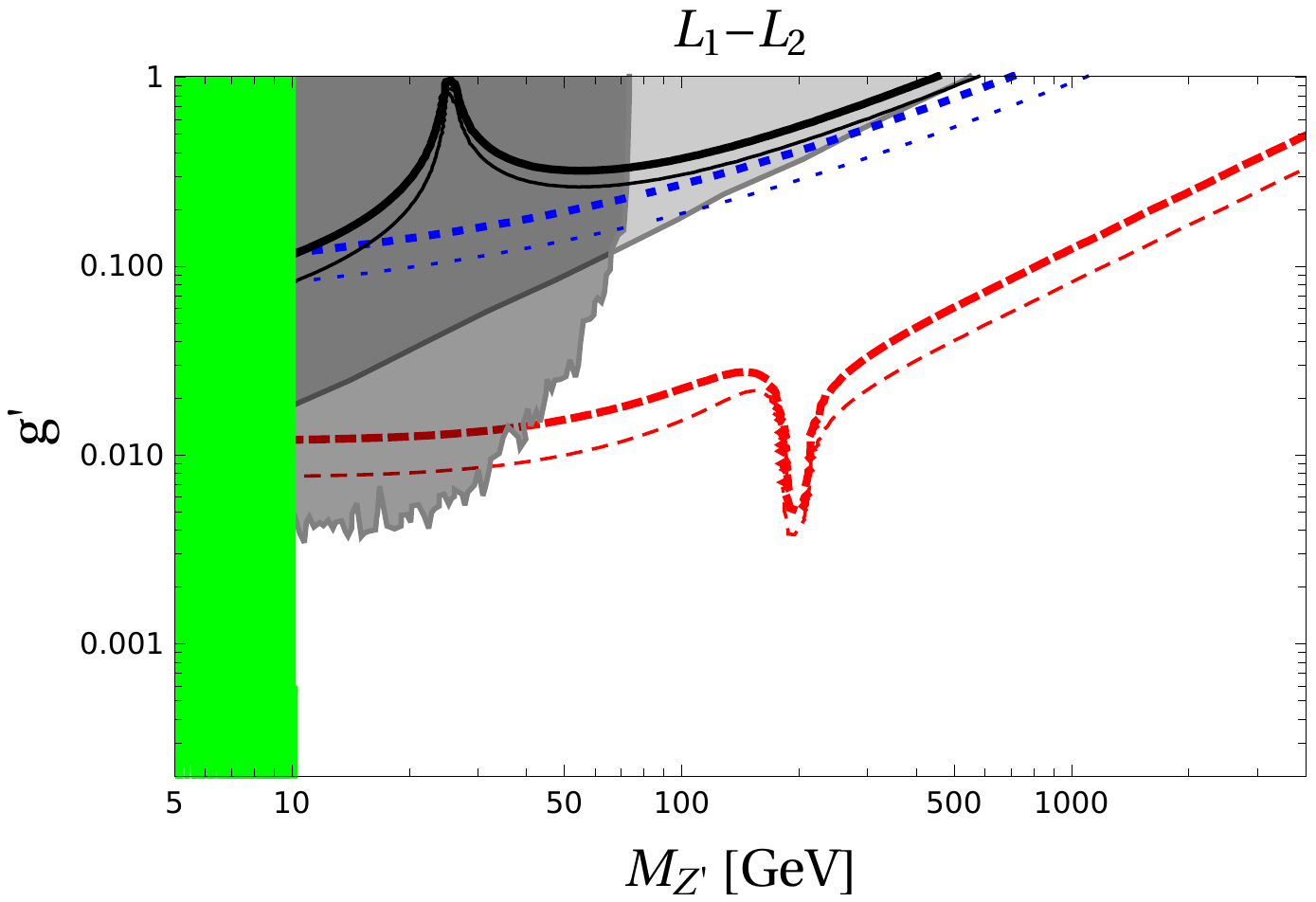}\quad
\includegraphics[scale=0.5]{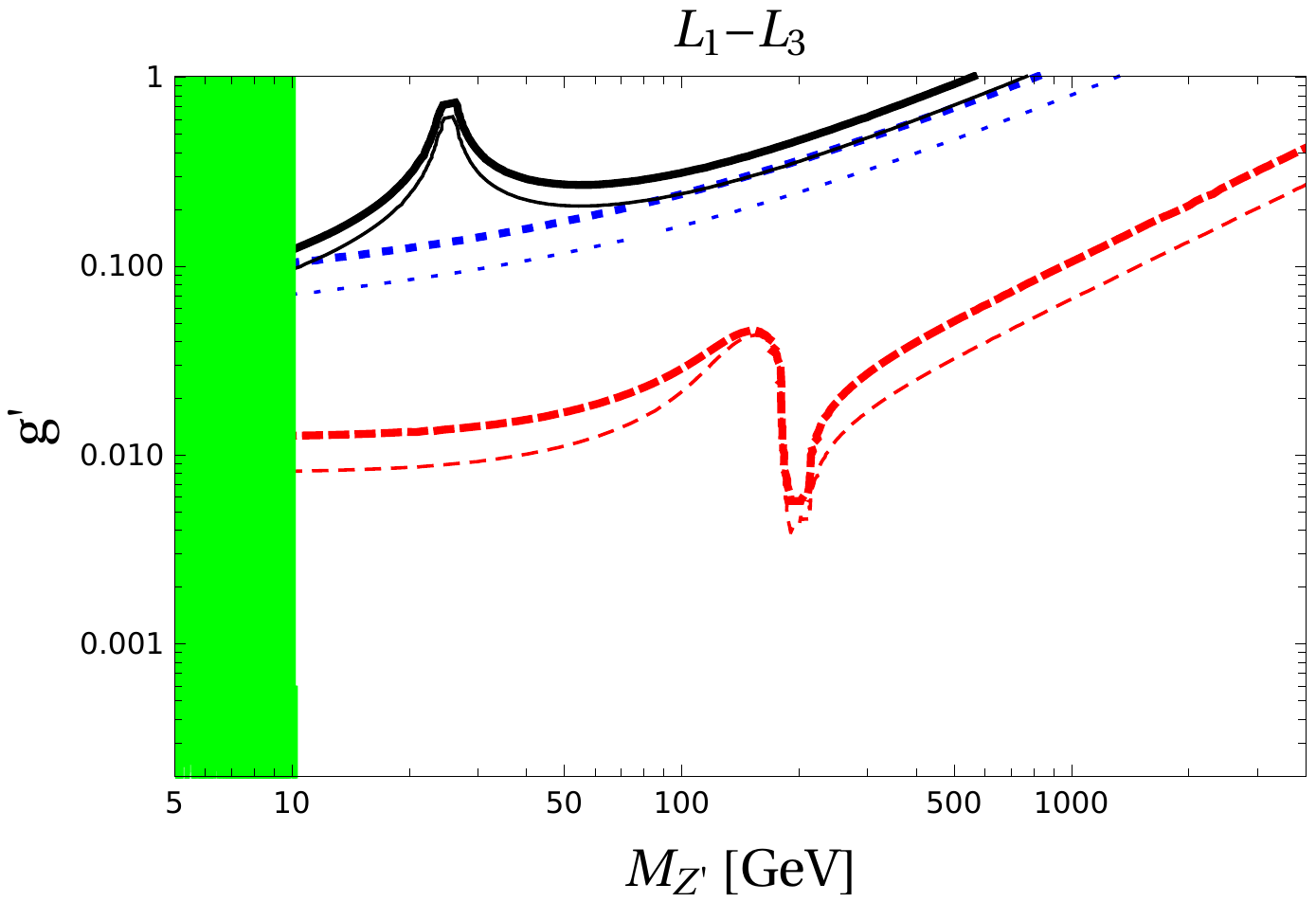} \\
\includegraphics[scale=0.5]{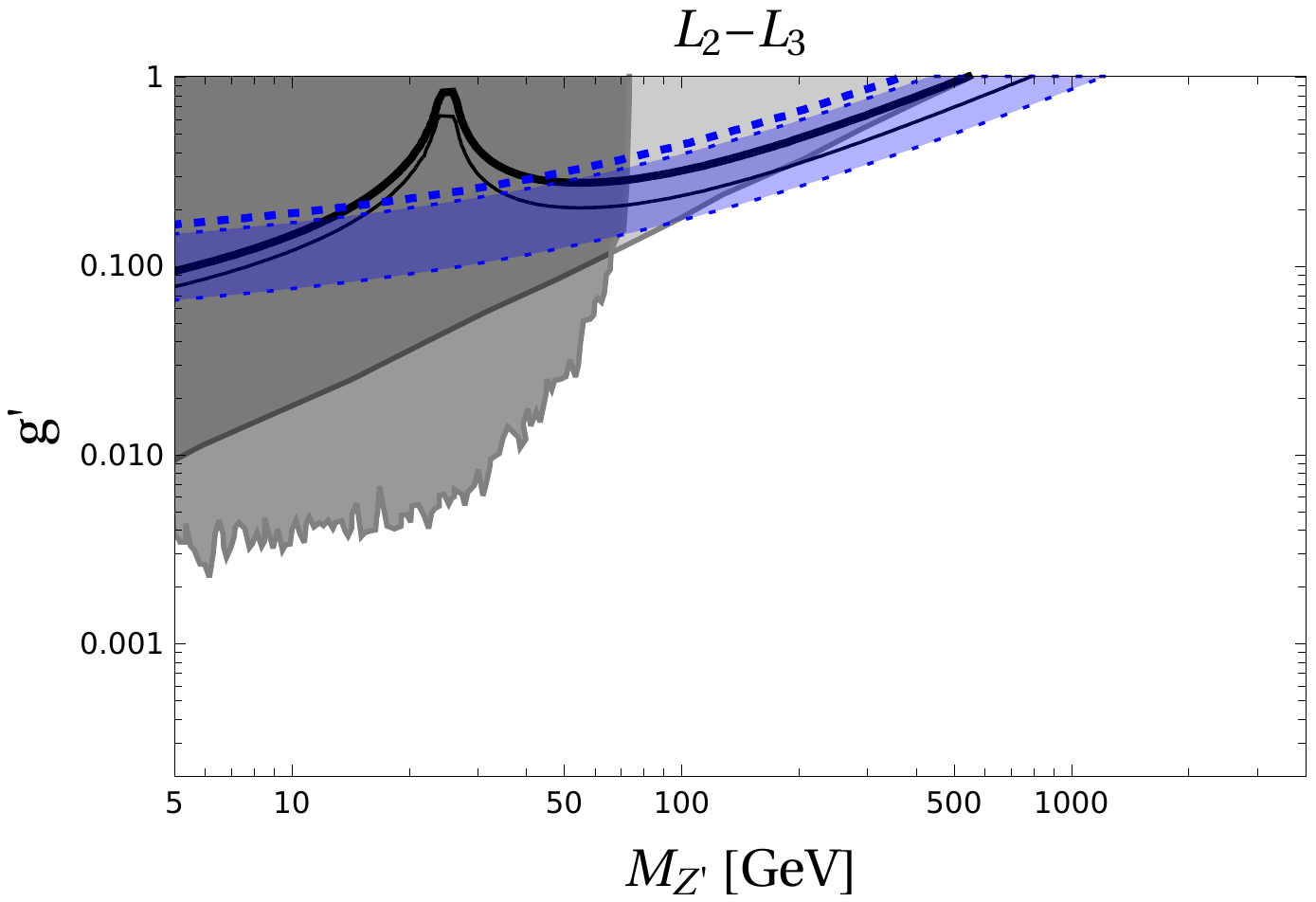}
\caption{
Bounds on $g^{\prime}$ and $M_{Z^{\prime}}$ for the $L_i-L_j$ models. The thin (thick) dashed-red, dotted-blue, and solid-black curves correspond to 1$\sigma$ (2$\sigma$) bounds from LEP search \cite{Schael:2013ita}, HFAG lepton universality test \cite{hfag16}, and SLD/LEP $Z$-decay lepton universality test \cite{Zll}, respectively.
The green region is excluded by BaBar bounds \cite{babar}.
The lighter- and darker-grey regions are excluded by neutrino-trident bound \cite{Mishra:1991bv, Altmannshofer:2014pba} and the LHC bound for $Z \rightarrow 4\mu$ \cite{Sirunyan:2018nnz}  respectively.
In the $L_2 - L_3$ case, the blue-shaded region is the 1$\sigma$-allowed region by HFAG lepton universality test. The 2$\sigma$-allowed region is below  the thick dotted-blue curve.
Note the absence of the BaBar and LEP bounds in this case where $Z'$ does not couple to electrons.
}
\label{LimLj}
\end{center}
\end{figure*}

In the $L_i-L_j$ case, the combined limits are shown in Fig.~\ref{LimLj}.
Bounds on $g^{\prime}$ and $M_{Z^{\prime}}$ for the $L_i-L_j$ models come from the LEP search \cite{Schael:2013ita}, HFAG lepton universality test \cite{hfag16}, and SLD/LEP $Z$-decay lepton universality test \cite{Zll}, respectively represented in Fig.~\ref{LimLj} as dashed red, dotted blue, and solid black curves.
On top of these constraints we also project the BaBar bounds \cite{babar} (green regions), neutrino-trident bound \cite{Mishra:1991bv, Altmannshofer:2014pba} (lighter-grey regions), and LHC bound for $Z \rightarrow 4\mu$ \cite{Sirunyan:2018nnz}. (darker-grey regions).
In the $L_2 - L_3$ case, the new gauge boson $Z'$ does not interact with electrons. Therefore there are no constraints from BaBar and LEP bounds.
Note that, when $L_1$ is involved, the LEP bounds on $e^{-}e^{+} \rightarrow \ell^{-}\ell^{+}$ processes give the most stringent bounds.
On the other hand, when $L_2$ is involved, the LHC bound for $Z \rightarrow 4\mu$ gives the strongest bound in the small mass region $M_{Z^{\prime}} \lesssim 70$ GeV at $2\sigma$.

\begin{figure*}[t]
\begin{center}
\includegraphics[scale=0.5]{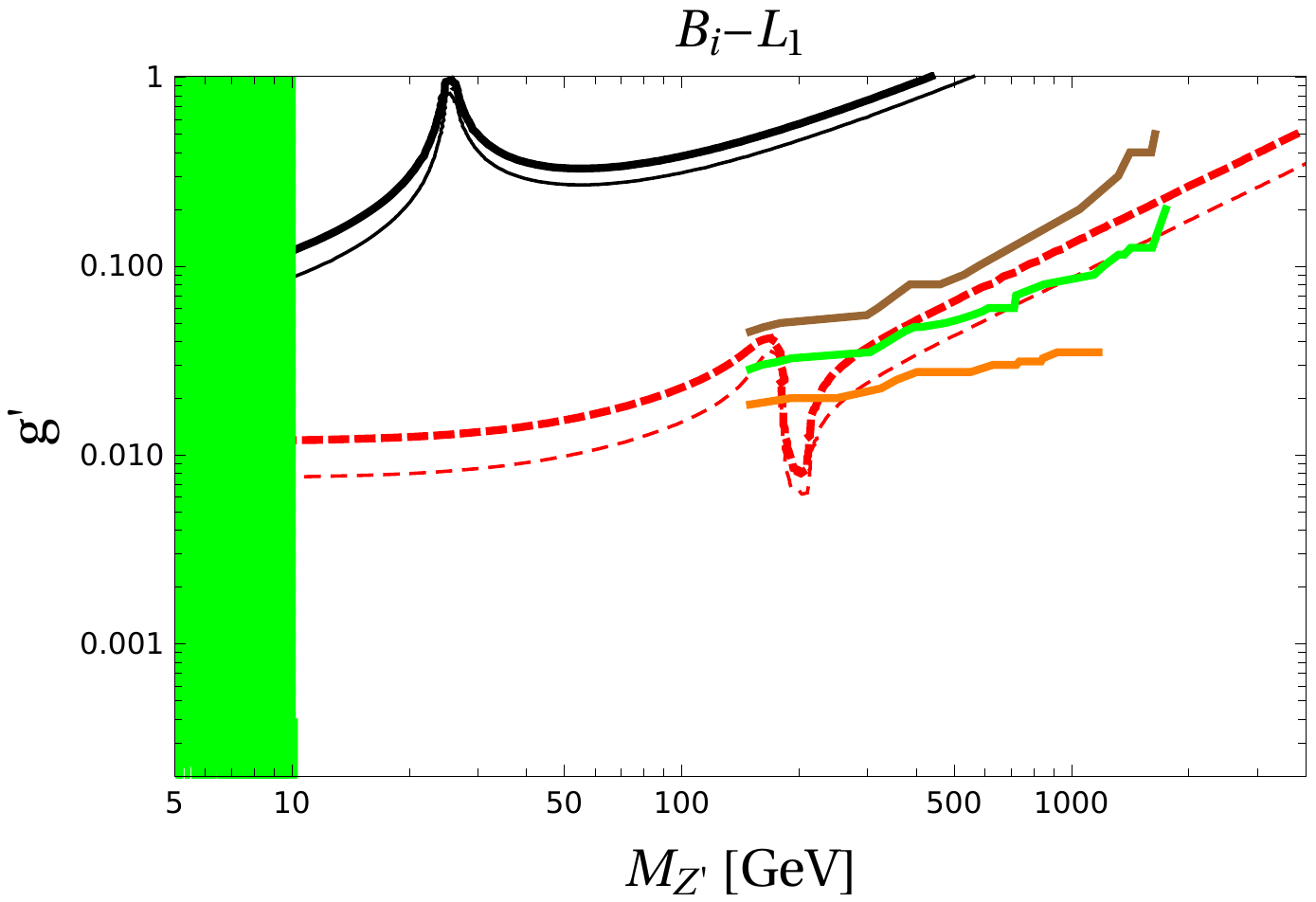}\quad
\includegraphics[scale=0.5]{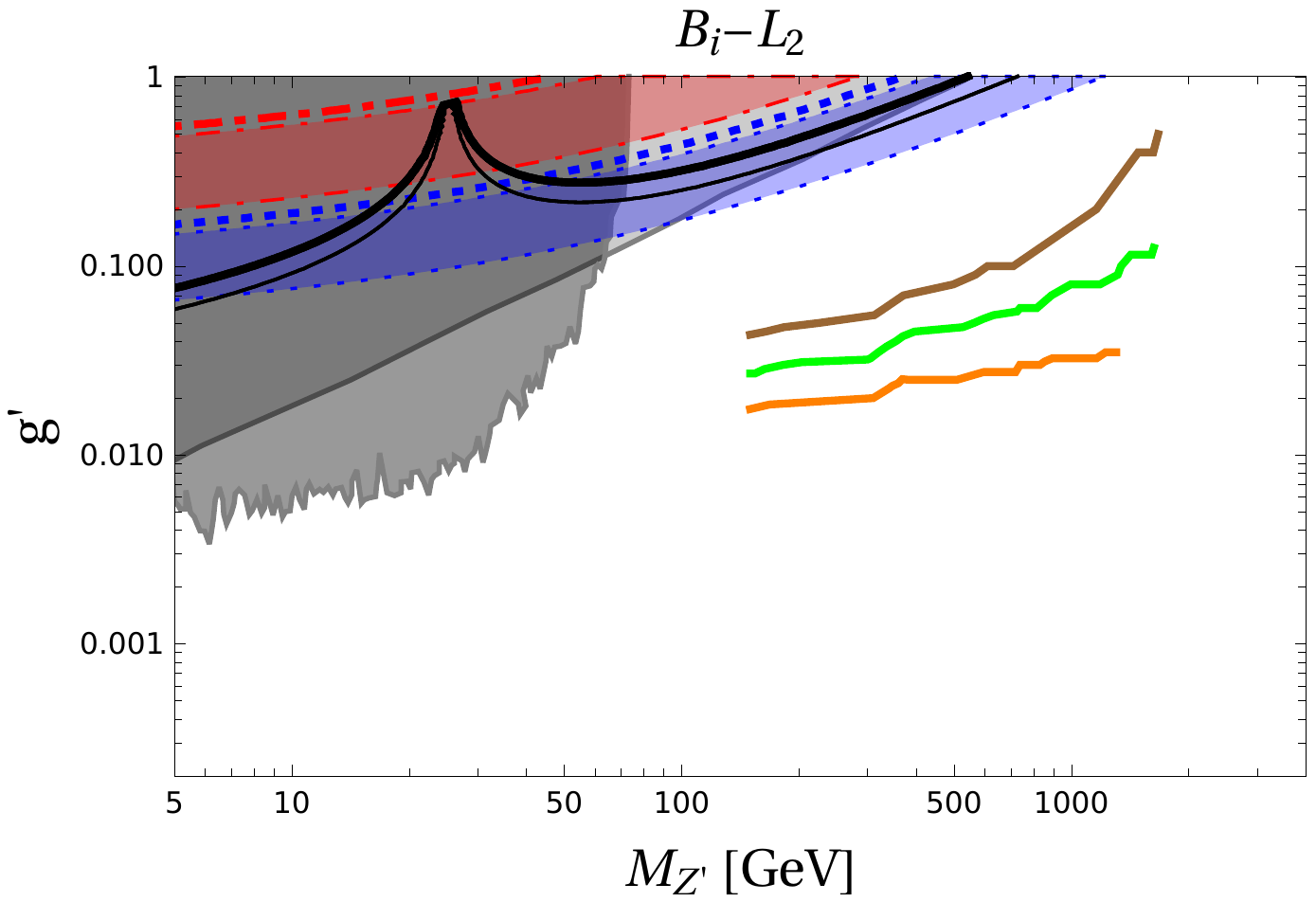}\\
\includegraphics[scale=0.5]{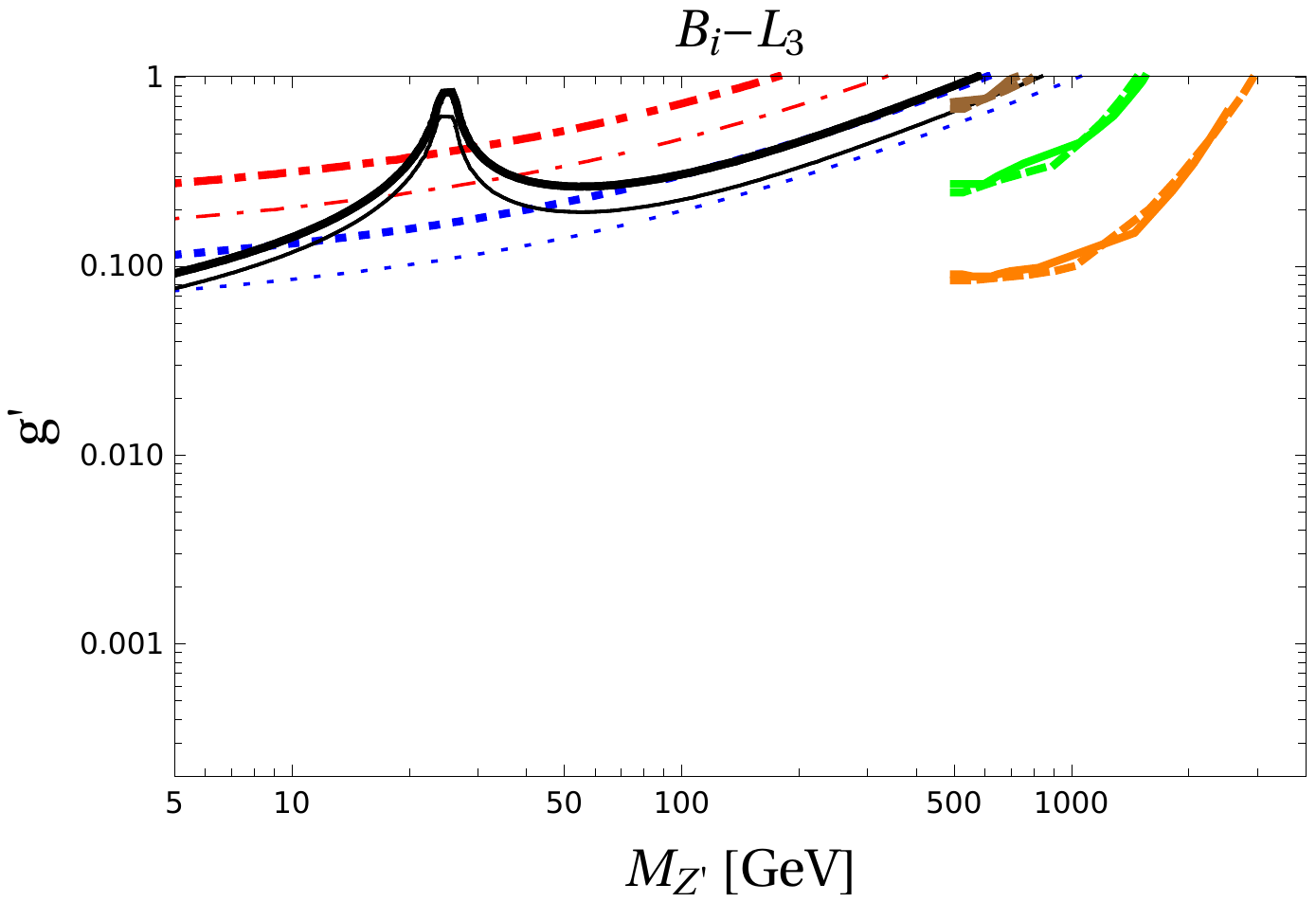}
\caption{
Bounds on $g^{\prime}$ and $M_{Z^{\prime}}$ for the $B_i-L_j$ models.
The green region is excluded by BaBar bounds \cite{babar}.
The solid(dashed)-orange, -green, and -brown curves correspond to the ATLAS (CMS) bounds for $B_{1}$, $B_{2}$, and $B_{3}$, respectively. Note that we considered only the bottom quark in the $B_{3}$ case.
The lighter- and darker-grey regions are excluded by neutrino-trident bound \cite{Mishra:1991bv, Altmannshofer:2014pba} and the LHC bound for $Z \rightarrow 4\mu$ \cite{Sirunyan:2018nnz}. 
{\it Upper left panel}: Bounds on $g^{\prime}$ and $M_{Z^{\prime}}$ for the $B_{i} - L_{1}$ models. The thin (thick) dashed-red and solid-black curves correspond to 1$\sigma$ (2$\sigma$) bounds from LEP search \cite{Schael:2013ita} and SLD/LEP $Z$-decay lepton universality test \cite{Zll}, respectively.
{\it Upper right panel}: Bounds on $g^{\prime}$ and $M_{Z^{\prime}}$ for the $B_{i} - L_{2}$ models.
The thin (thick) solid-black curve corresponds to 1$\sigma$ (2$\sigma$) bounds from the SLD/LEP $Z$-decay lepton universality test \cite{Zll}. The blue (red) shaded region between thin dotted-blue (dot-dashed-red) curves is the 1$\sigma$-allowed region by HFAG lepton universality test \cite{hfag16} for the $B_1$ ($B_2$) case. The corresponding 2$\sigma$-allowed region is below  the thick dotted-blue (dot-dashed-red) curve.
{\it Lower panel}: Bounds on $g^{\prime}$ and $M_{Z^{\prime}}$ for the $B_{i} - L_{3}$ models.
The thin (thick) solid-black curve corresponds to 1$\sigma$ (2$\sigma$) bound from the SLD/LEP $Z$-decay lepton universality test \cite{Zll}. The thin (thick) dot-dashed-red, dotted-blue curves correspond to 1$\sigma$ (2$\sigma$) bounds from the HFAG lepton universality test \cite{hfag16} for the $B_{2}$ and $B_1$ cases,  respectively.
}
\label{BimLj}
\end{center}
\end{figure*}

In the $B_i-L_j$ case, the combined limits are shown in Fig.~\ref{BimLj}.
In the $B_i - L_2$ and $B_i - L_3$ cases, the new gauge boson $Z'$ does not interact with electrons. Therefore there are no constraints from BaBar and LEP bounds.
When $L_1$ is involved, the LEP bounds on $e^{-}e^{+} \rightarrow \ell^{-}\ell^{+}$ processes give the most stringent bounds, except for $300$ GeV $\lesssim M_{Z^{\prime}} \lesssim 1.5$ TeV, where the LHC bounds give the severe constraints.
For $B_i - L_2$ cases, the LHC bound for $Z \rightarrow 4\mu$ gives the strongest bound in the small mass region $M_{Z^{\prime}} \lesssim 70$ GeV at $2\sigma$.

\begin{figure*}[t]
\begin{center}
\includegraphics[scale=0.5]{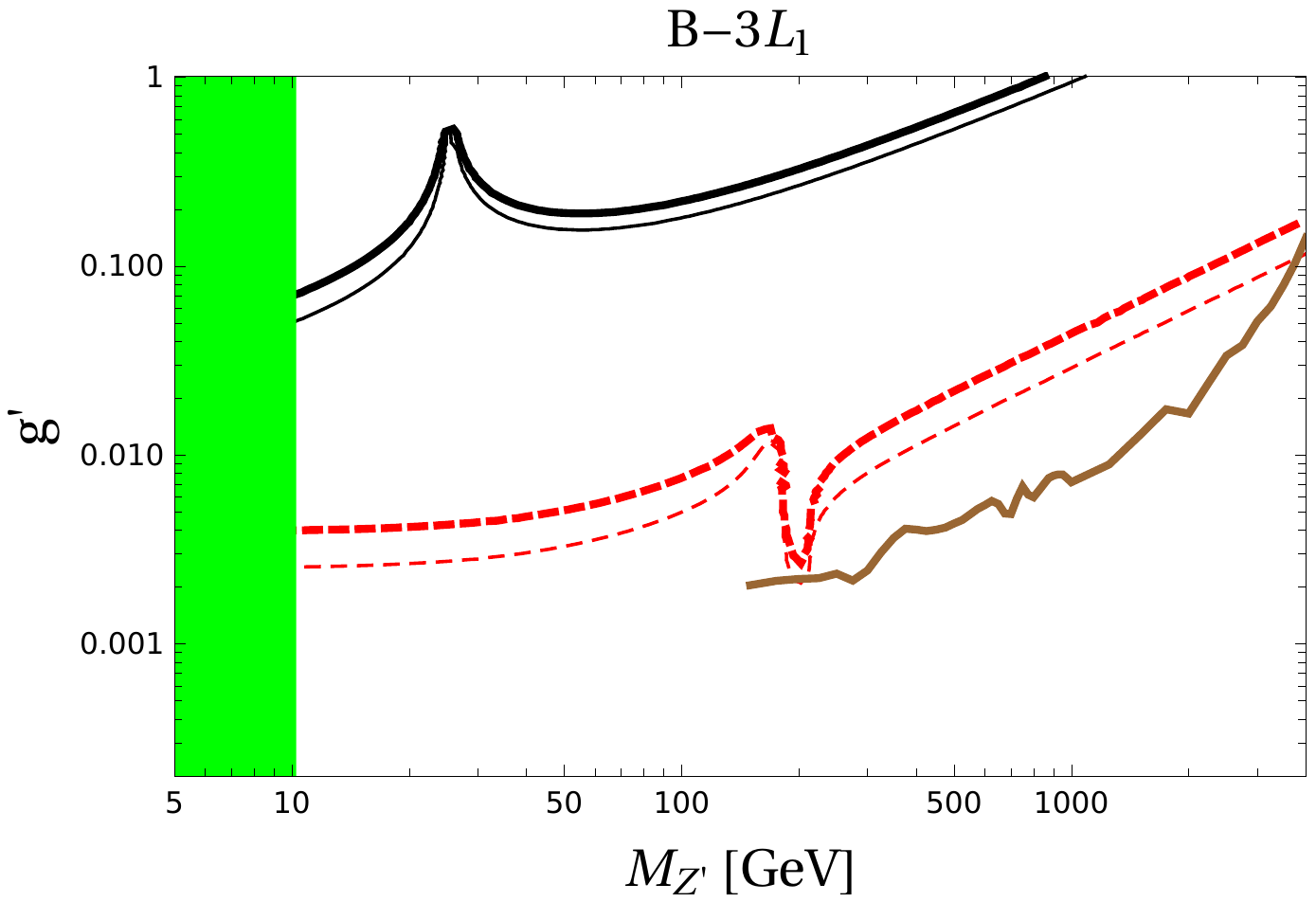}\quad
\includegraphics[scale=0.5]{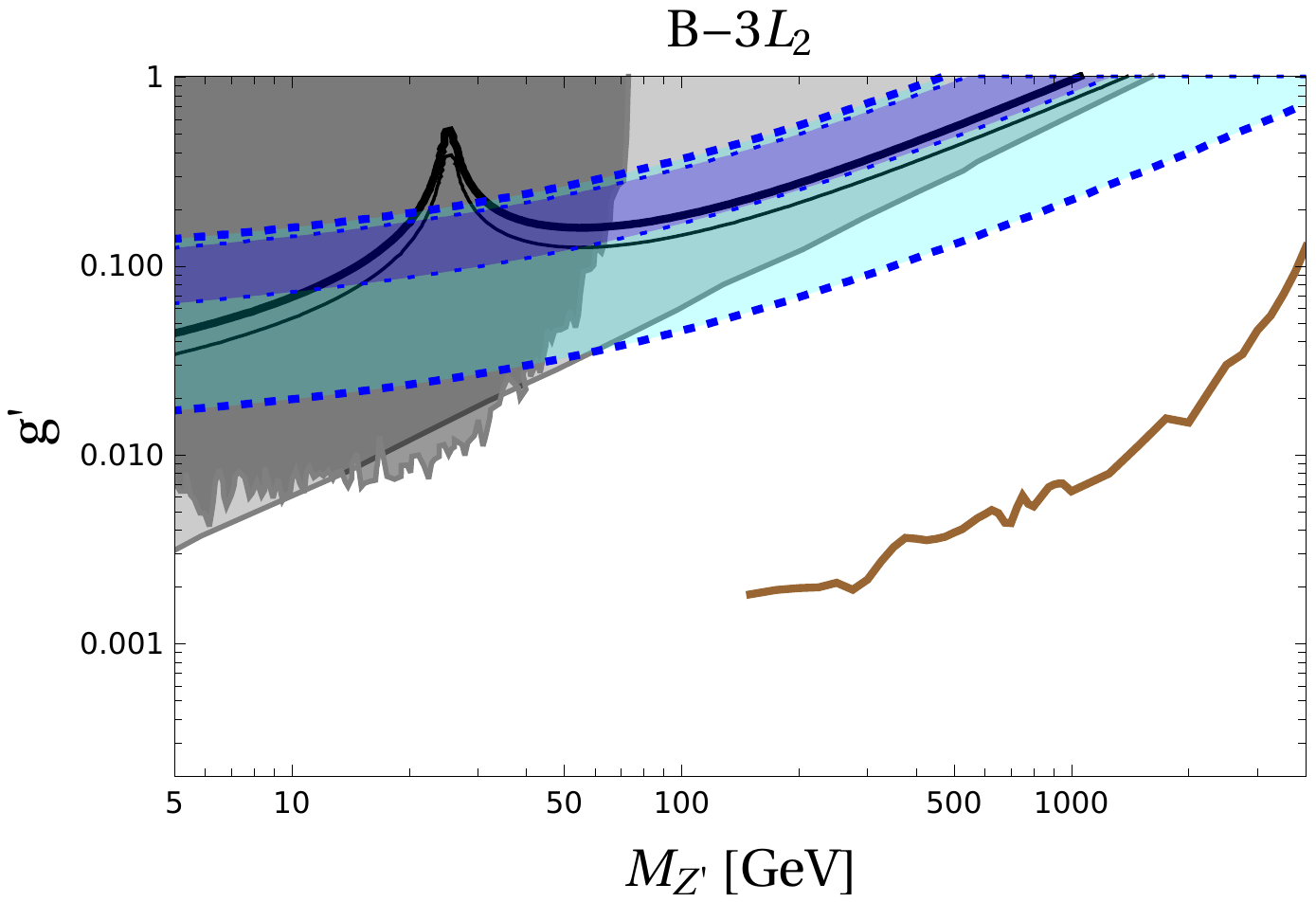}\\
\includegraphics[scale=0.5]{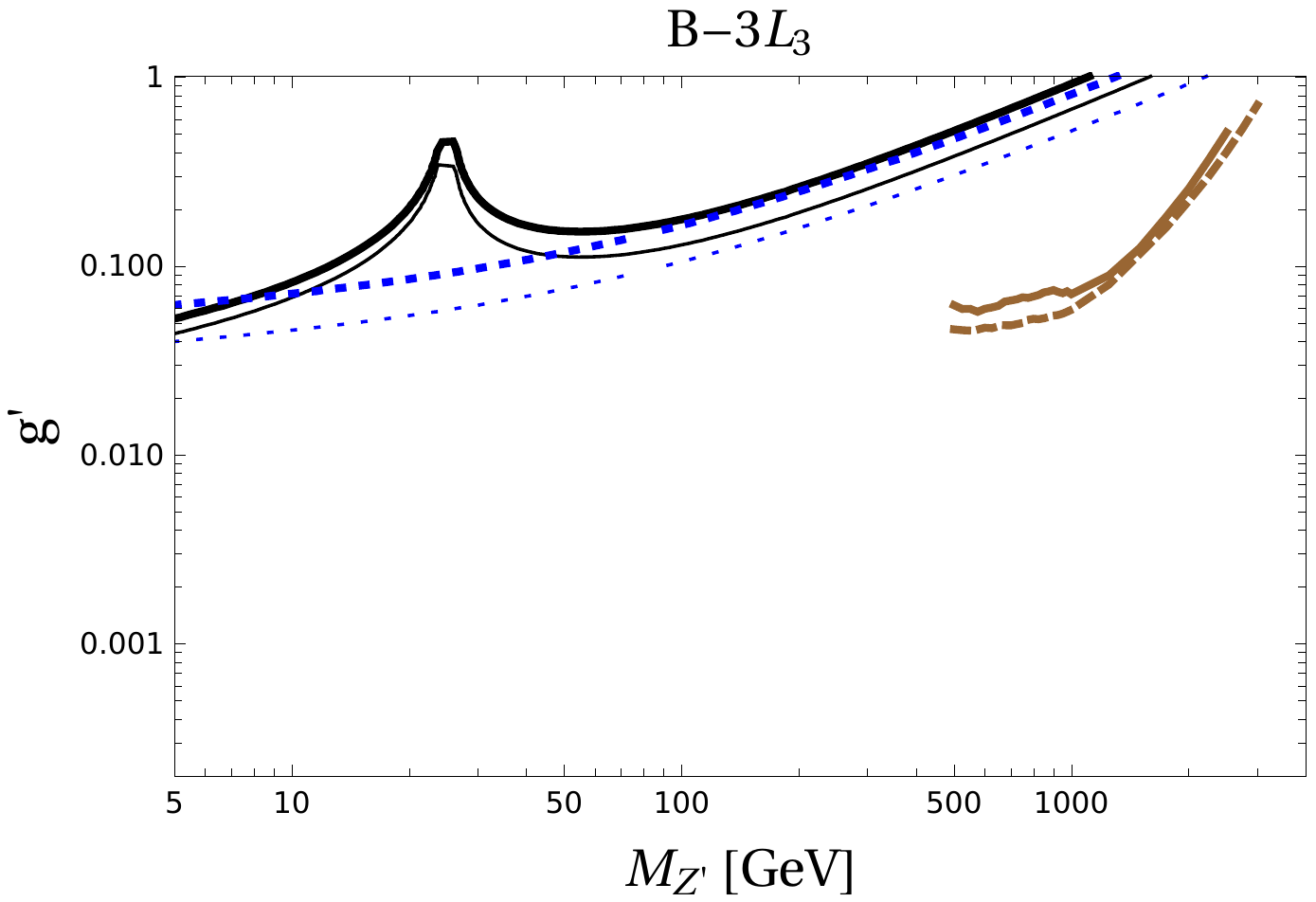}
\caption{
{\it Upper left panel}: Bounds on $g^{\prime}$ and $M_{Z^{\prime}}$ for the $B - 3L_{1}$ model. The thin (thick) dashed-red and solid-black curves correspond to 1$\sigma$ (2$\sigma$) bounds from LEP search \cite{Schael:2013ita} and SLD/LEP $Z$-decay lepton universality test \cite{Zll}, respectively.
The green region is excluded by BaBar bounds \cite{babar}. The brown curve is the LHC bound from ATLAS observed limits.
{\it Upper right panel}: Bounds on $g^{\prime}$ and $M_{Z^{\prime}}$ for the $B - 3L_{2}$ model.
The thin (thick) solid-black curve corresponds to 1$\sigma$ (2$\sigma$) bound from the SLD/LEP $Z$-decay lepton universality test \cite{Zll}.  The blue (cyan) shaded region between thin (thick) dotted-blue curves is the 1$\sigma$ (2$\sigma$) allowed region by the HFAG lepton universality test \cite{hfag16}.
The brown curve is the LHC bound from ATLAS observed limits.
The lighter- and darker-grey regions are excluded by neutrino-trident bound \cite{Mishra:1991bv, Altmannshofer:2014pba} and the LHC bound for $Z \rightarrow 4\mu$ \cite{Sirunyan:2018nnz}.
Therefore, only a small region $60$ GeV $\lesssim M_{Z^{\prime}} \lesssim 150$ GeV is consistent with all the constraints at $2\sigma$.
{\it Lower panel}: Bounds on $g^{\prime}$ and $M_{Z^{\prime}}$ for the $B - 3L_{3}$ model. The thin (thick) dotted blue and solid black curves correspond to 1$\sigma$ (2$\sigma$) bounds from the HFAG lepton universality test \cite{hfag16} and SLD/LEP $Z$-decay lepton universality test \cite{Zll}, respectively.
The solid and dashed brown curves are the LHC bounds from ATLAS and CMS respectively.
}
\label{Bm3Li}
\end{center}
\end{figure*}

In the $B-3L_i$ case, the combined limits are shown in Fig.~\ref{Bm3Li}.
For all $B - 3L_{i}$ cases, the LHC bounds become the strongest bounds for large $M_{Z^{\prime}}$ region; $M_{Z^{\prime}} \gtrsim 150\;{\rm GeV}$ for $i=1,2$ cases and $M_{Z^{\prime}} \gtrsim 500\;{\rm GeV}$ for $i=3$ case. In the $B - 3L_{1}$ case, the LEP bounds on $e^{-}e^{+} \rightarrow \ell^{-}\ell^{+}$ processes give the most stringent bounds on small $M_{Z^{\prime}}$ region, $10\;{\rm GeV} \lesssim M_{Z^{\prime}} \lesssim 150\;{\rm GeV}$. The $B - 3L_{2}$ case is particularly interesting. Due to the strong bounds from HFAG lepton universality test, only a small region $60\;{\rm GeV} \lesssim M_{Z^{\prime}} \lesssim 150\;{\rm GeV}$ is consistent with all the constraints at $2\sigma$.

\section{Conclusions}
\label{sec:con}
In this paper we studied three types of anomaly-free flavored gauge symmetries, namely $L_{i} - L_{j}$, $B_{i} - L_{j}$, and $B - 3L_{i}$, where $i,j=1,2,3$ are flavor indices.
Utilising the lepton universality test in the $Z$ and $\tau/\mu$ decays, LEP, BaBar and LHC searches, and neutrino trident and $Z\rightarrow 4\mu$ production searches, we investigated phenomenological implications of the flavored gauge boson $Z^{\prime}$ and put various constraints.

The combined limits for $L_i - L_j$, $B_i - L_j$, and $B - 3L_i$ scenarios are shown respectively in Fig.~\ref{LimLj}, Fig.~\ref{BimLj}, and Fig.~\ref{Bm3Li}.
When $L_1$ is involved, the LEP bounds on the $e^{-}e^{+} \rightarrow \ell^{-}\ell^{+}$ processes give the most stringent bounds in most parameter regions.
When the quark sector is included the LHC bound becomes the strongest constraints in large $M_{Z^{\prime}}$ region.
On the other hand, when $L_2$ is involved, the bounds from neutrino trident and $Z \rightarrow 4\mu$ productions give the strongest constraints in small $M_{Z^{\prime}}$ region. Where these bounds are not applicable, the lepton universality test puts a mild limit.
Interestingly, in the $B - 3L_2$ case, the slight deviation from lepton universality reported by HFAG limits the parameter space so strongly that only a small region around  $M_{Z^{\prime}}\sim 100$ GeV and $g'\sim 0.06$ remains consistent with all the constraints at $2\sigma$ level.

\acknowledgments{
We thank Hongkai Liu for pointing out our mistakes on the LHC bounds for the $B-3L_i$ models.
The work of Jinsu Kim was supported by Alexander von Humboldt Foundation.
Jongkuk Kim was supported by the National Research Foundation of Korea(NRF) grant funded by the Korea government(MEST) (NRF-2016R1A2B4012302,  NRF-2018R1D1A1B07051127).
}

\begin{appendix}

\section{Expressions for $\delta_{\ell\ell}$}
The expressions for $\delta_{\ell\ell}$ introduced in Section \ref{sec:LUtest} for all the combinations we considered are summarized in Table \ref{tab:deltaellell}. They are calculated at one-loop level.
\begin{table}[t]
\centering\small
\begin{tabular}{ c | c c c | c c c | c c c}
	& $L_{1} - L_{2}$ & $L_{1} - L_{3}$ & $L_{2} - L_{3}$
	& $B_{i} - L_{1}$ & $B_{i} - L_{2}$ & $B_{i} - L_{3}$
	& $B - 3L_{1}$ & $B - 3L_{2}$ & $B - 3L_{3}$ \\
	\hline
	$\delta_{\mu\mu}$ & 0 & $-\widetilde{\delta}$ & $\widetilde{\delta}$
	& $-\widetilde{\delta}$ & $\widetilde{\delta}$ & 0
	& $-3\widetilde{\delta}$ & $3\widetilde{\delta}$ & 0  \\
    $\delta_{\tau\tau}$ & $-\widetilde{\delta}$ & 0 & $\widetilde{\delta}$
    & $-\widetilde{\delta}$ & 0 & $\widetilde{\delta}$
    & $-3\widetilde{\delta}$ & 0 & $3\widetilde{\delta}$
\end{tabular}
\caption{Expressions for $\delta_{\ell\ell} \equiv (\Gamma_{Z\rightarrow\ell^{+}\ell^{-}}/\Gamma_{Z\rightarrow e^{+}e^{-}}) -1$ at one-loop level.}
\label{tab:deltaellell}
\end{table}
The quantity $\widetilde{\delta}$ is defined by
\begin{align}
	\widetilde{\delta} \equiv
	\frac{2g_{L}^{2}{\rm Re}(\delta g_{L}^{e})}
	{(g_{L}^{e})^{2}+(g_{R}^{e})^{2}}\,,
\end{align}
where
\begin{align}
	\delta g_{L}^{e} =
	\frac{(g^{\prime})^{2}}{8\pi^{2}}K(r)\,,
\end{align}
with $r \equiv M_{Z^{\prime}}^{2}/M_{Z}^{2}$, $g_{L}^{e} \approx -0.27$ and $g_{R}^{e} \approx 0.23$. The real part of $K(r)$, which is the only relevant quantity, is given by
\begin{align}
	{\rm Re}K(r)
	&=-\frac{7}{2}-2r-2(1+r)^{2}\ln^{2}r
	+\ln r \left[
	2(1+r)^{2}\ln(1+r)-3-2r
	\right]
	\nonumber\\
	&\quad
	-2(1+r)^{2}{\rm Li}_{2}(-1/r)\,.
\end{align}

\section{Expressions for $\delta_{\ell/\ell^{\prime}}$ and $\delta_{\pi,K}$}
The expressions for $\delta_{\ell/\ell^{\prime}}$ and $\delta_{\pi,K}$ for all combinations we considered are summarized in Tables \ref{tab:deltamutau1} and \ref{tab:deltamutau2}. They are calculated at one-loop level.
\begin{table}[t]
\centering
\begin{tabular}{ c | c c c | c c c }
	& $L_{1} - L_{2}$ & $L_{1} - L_{3}$ & $L_{2} - L_{3}$
	& $B - 3L_{1}$ & $B - 3L_{2}$ & $B - 3L_{3}$ \\
	\hline
	$\delta_{\tau/\mu}$ & $-\delta$ & $\delta$ & 0
	& 0 & 0 & 0 \\
	$\delta_{\tau/e}$ & $-\delta$ & 0 & $\delta$
	& 0 & 0 & 0 \\
    $\delta_{\mu/e}$ & 0 & $-\delta$ & $\delta$
    & 0 & 0 & 0    \\
    $\delta_{\pi}$ & $-\delta$ & $\delta$ & 0
    & 0 & $-3\delta$ & $3\delta$ \\
    $\delta_{K}$ & $-\delta$ & $\delta$ & 0
    & 0 & $-3\delta$ & $3\delta$
\end{tabular}
\caption{Expressions for $\delta_{\ell/\ell^{\prime}} \equiv (g_{\ell}/g_{\ell^{\prime}})-1$ and $\delta_{\pi,K} \equiv (g_{\tau}/g_{\mu})_{\pi,K} -1$.}
\label{tab:deltamutau1}
\end{table}
\begin{table}[t]
\centering
\begin{tabular}{ c | c c c c c c c }
	& $B_{i} - L_{1}$ & $B_{1} - L_{2}$ & $B_{2} - L_{2}$ & $B_{3}-L_{2}$ & $B_{1}-L_{3}$ & $B_{2}-L_{3}$ & $B_{3}-L_{3}$ \\
	\hline
	$\delta_{\tau/\mu}$ & 0 & 0 & 0 & 0 & 0 & 0 & 0 \\
	$\delta_{\tau/e}$ & 0 & 0 & 0 & 0 & 0 & 0 & 0 \\
	$\delta_{\mu/e}$ & 0 & 0 & 0 & 0 & 0 & 0 & 0 \\
	$\delta_{\pi}$ & 0 & $-\delta$ & 0 & 0 & $\delta$ & 0 & 0 \\
	$\delta_{K}$ & 0 & $-\delta/2$ & $-\delta/2$ & 0 & $\delta/2$ & $\delta/2$ & 0
\end{tabular}
\caption{Expressions for $\delta_{\ell/\ell^{\prime}} \equiv (g_{\ell}/g_{\ell^{\prime}})-1$ and $\delta_{\pi,K} \equiv (g_{\tau}/g_{\mu})_{\pi,K} -1$.}
\label{tab:deltamutau2}
\end{table}
The quantity $\delta$ is defined by
\begin{align}
	\delta \equiv
	\frac{6(g^{\prime})^{2}}{16\pi^{2}}
	\frac{\ln(M_{W}^{2}/M_{Z^{\prime}}^{2})}
	{1-M_{Z^{\prime}}^{2}/M_{W}^{2}}\,.
\end{align}

\section{Expressions for differential cross sections for $e^{+}e^{-} \rightarrow \ell^{+}\ell^{-}$}
The differential cross section for $e^{+}e^{-} \rightarrow e^{+}e^{-}$ at tree level is given by
\begin{align}
	\frac{d\sigma}{d\Omega}
	= \frac{1}{64\pi^{2}s}\big[
	32\pi^{2}s^{2}A^{2}
	+16\pi^{2}u^{2}(B^{2}+C^{2})
	+8\pi^{2}t^{2}D^{2}
	\big]\,,
\end{align}
where
\begin{align}
	A &=
	\frac{\alpha(t_{W}^{2} - 1)}{2(t-M_{Z}^{2})}
	+\frac{\alpha^{\prime}}{t-M_{Z^{\prime}}^{2}}
	+\frac{\alpha}{t}\,,\\
	B &=
	\alpha t_{W}^{2}\left(
	\frac{1}{s-M_{Z}^{2}}
	+\frac{1}{t-M_{Z}^{2}}
	\right)
	+\alpha^{\prime} \left(
	\frac{1}{s-M_{Z^{\prime}}^{2}}
	+\frac{1}{t-M_{Z^{\prime}}^{2}}
	\right)
	+\alpha \frac{s+t}{st}\,,\\
	C &=
	\alpha\frac{(t_{W}^{2}-1)^{2}}{4t_{W}^{2}}
	\left(
	\frac{1}{s-M_{Z}^{2}}
	+\frac{1}{t-M_{Z}^{2}}
	\right)
	+\alpha^{\prime} \left(
	\frac{1}{s-M_{Z^{\prime}}^{2}}
	+\frac{1}{t-M_{Z^{\prime}}^{2}}
	\right)
	+\alpha \frac{s+t}{st}\,,\\
	D &=
	\alpha\frac{s-2M_{Z}^{2}}{s(s-M_{Z}^{2})}
	+\frac{2\alpha^{\prime}}{s-M_{Z^{\prime}}^{2}}
	+\frac{\alpha t_{W}^{2}}{s-M_{Z}^{2}} \,.
\end{align}
Here we neglected small electron mass and defined $\alpha^{\prime} \equiv (g^{\prime})^{2}/4\pi$ and $t_{W} \equiv \sin\theta_{W}/\cos\theta_{W}$ with $\theta_{W}$ being the Weinberg angle. We also introduced the Mandelstam variables $s$, $t$, and $u$, which are given by
\begin{align}
	t = -\frac{s}{2}(1-\cos\theta)\,,\qquad
	u = -\frac{s}{2}(1+\cos\theta)\,,
\end{align}
with $\theta$ being the angle between the initial electron and the final electron.

The differential cross section for $e^{+}e^{-} \rightarrow \ell^{+}\ell^{-}$ ($\ell\neq e$) at tree level is given by
\begin{align}
	\frac{d\sigma}{d\Omega}
	= \frac{1}{64\pi^{2}s}\big[
	16\pi^{2}u^{2}(E^{2}+F^{2})
	+8\pi^{2}t^{2}G^{2}
	\big]\,,
\end{align}
where
\begin{align}
	E &=
	\frac{\alpha t_{W}^{2}}{s-M_{Z}^{2}}
	-\frac{\alpha^{\prime}}{s-M_{Z^{\prime}}^{2}}
	+\frac{\alpha}{s}\,,\\
	F &=
	\frac{(t_{W}^{2}-1)^{2}}{4t_{W}^{2}}
	\frac{\alpha}{s-M_{Z}^{2}}
	-\frac{\alpha^{\prime}}{s-M_{Z^{\prime}}^{2}}
	+\frac{\alpha}{s}\,,\\
	G &=
	\alpha\frac{s-2M_{Z}^{2}}{s(s-M_{Z}^{2})}
	-\frac{2\alpha^{\prime}}{s-M_{Z^{\prime}}^{2}}
	+\frac{\alpha t_{W}^{2}}{s-M_{Z}^{2}} \,.
\end{align}
Here we again neglected small final lepton state masses.

\end{appendix}



\end{document}